\documentclass[11pt,a4paper]{article}

\usepackage{amsmath,amsfonts,amssymb,amsthm,cite}

\evensidemargin=\oddsidemargin
\advance\topmargin by -\headheight
\advance\topmargin by -\headsep

\newtheorem{thm}{Theorem}[section]
\newtheorem{lem}[thm]{Lemma}

\theoremstyle{definition}                    
\newtheorem{defn}[thm]{Definition}
\theoremstyle{remark}
\newtheorem{rem}[thm]{Remark}

\newcommand{\A}{\mathcal{A}}        

\DeclareMathOperator{\Diff}{Diff}

\newcommand{\B}{\mathcal{B}}        
\newcommand{\C}{\mathbb{C}}
\newcommand{\CC}{\mathcal{C}^\infty}     
\newcommand{\D}{\mathcal{D}}        
\newcommand{\Dslash}{{D\mkern-11.5mu/\,}} 
\newcommand{\dslash}{{\partial\mkern-10mu/\,}}
\newcommand{\eps}{\varepsilon}      
\newcommand{\ga}{\gamma}            
\renewcommand{\H}{\mathcal{H}}      

\newcommand{\K}{\mathcal{K}}
\renewcommand{\L}{\mathcal{L}}
\newcommand{\mop}{\mathbin{\star_{_\theta}}}
\newcommand{\Mop}{\mathbin{\star_{_\Theta}}}
\newcommand{\N}{\mathbb{N}}         
\newcommand{\Oh}{\mathcal{O}}       
\newcommand{\pa}{\partial}          
\newcommand{\R}{\mathbb{R}}         
\newcommand{\sepword}[1]{\quad\mbox{#1}\quad} 
\newcommand{\set}[1]{\{\,#1\,\}}     
\DeclareMathOperator{\spec}{Spect}
\renewcommand{\SS}{\mathcal{S}}     

\newcommand{\Th}{\Theta}
\renewcommand{\th}{\theta}          
\newcommand{\thalf}{\tfrac{1}{2}}   
\DeclareMathOperator{\Tr}{Tr}       

\DeclareMathOperator{\Aut}{Aut}
\DeclareMathOperator{\Int}{Int}
\DeclareMathOperator{\Out}{Out}     
\renewcommand{\.}{\cdot}            
\newcommand{\Z}{\mathbb{Z}}
\newcommand{\T}{\mathbb{T}}
\newcommand{\slim}{\mathop{\mathrm{s\mbox{-}lim}}}  
\newcommand{\hideqed}{\renewcommand{\qed}{}}        

\begin{document}

\thispagestyle{empty}

\begin{center}

CENTRE DE PHYSIQUE TH\'EORIQUE$\,^1$\\
CNRS--Luminy, Case 907\\
13288 Marseille Cedex 9\\
FRANCE\\

\vspace{3cm}

{\Large\textbf{The spectral action for Moyal planes}} \\
\vspace{1.5cm}

{\large Victor GAYRAL$\,^2$ and Bruno IOCHUM$\,^2$} \\

\vspace{3cm}

{\large\textbf{Abstract}}
\end{center}
\quad Extending a result of D. V. Vassilevich \cite{Vass}, we obtain the asymptotic expansion
for the trace of a spatially regularized heat operator $L^\Th(f)e^{-t \triangle^\Th}$, where
$\triangle^\Th$ is a generalized Laplacian defined with
Moyal products and $L^\Th(f)$ is Moyal left multiplication. The Moyal planes corresponding
to any skewsymmetric matrix $\Th$ being spectral triples \cite{Himalia}, the spectral action
introduced in noncommutative geometry by A. Chamseddine and A. Connes \cite{AliAlain2} is
computed. This result generalizes the Connes-Lott action
\cite{ConnesL1} previously computed by Gayral \cite{Gayral} for symplectic $\Th$.

\vspace{2cm}

CPT-2004/P.005\\
hep-th/0402147
\vspace{2.5cm}

\noindent $^1$ UMR 6207

- Unit\'e Mixte de Recherche du CNRS et des
Universit\'es Aix-Marseille I, Aix-Marseille II et de l'Universit\'e
du Sud Toulon-Var

- Laboratoire affili\'e \`a la FRUMAM - FR 2291

\noindent $^2$ Also at Universit\'e de Provence,

 gayral@cpt.univ-mrs.fr, iochum@cpt.univ-mrs.fr

\newpage

\section{Introduction}

Since few years, the interest in noncommutative field theory has been
renewed in many works. The noncommutative space is quite
often of Moyal type, involving either noncommutative tori or Moyal planes
(see \cite{KotSch,DougNek} for recent reviews).

Historically, Moyal \cite{Moyal} has
tried to build quantum mechanics with a statistical point of view
using a phase space approach. Actually, his idea was that
the formalism of quantum theory allows to derive the phase space
distributions $F(p,q)$ when a theory of functions of
noncommuting observables is specified and
conversely. This type of consideration was initiated by Wigner
\cite{Wigner} with a formula for $F(p,q)$
using Fourier transform and for canonical conjugate coordinates and momenta
by Weyl \cite{Weyl} with group theoretical motivations.
A noncommutative star product was previously explicitly
given by Groenewold \cite{Groen}. In fact, the use of quantification by deformation
\cite{Bayenetall,RieffelBook} has been intensively
investigated since it yields a continuous path between classical
and quantum mechanics. In the meantime, the Weyl--Wigner quantification
process was also interesting for the pseudodifferential operators theory \cite{Folland}.
The Seiberg--Witten \cite{SeibWit} map allows to go from ordinary
to noncommutative gauge field theory and the replacement of the
ordinary commutative product of functions by the Moyal noncommutative one is
now ubiquitous in string theory where the effective low energy theory of
D-branes with B-field background lives in a noncommutative space.

The mathematical background of these different developments for quantization within
noncommutative geometry \cite{Book,Polaris} where noncommutative tori
\cite{Connes80,Rieffel} play an important role \cite{ConDougSch},
includes the construction of new
spectral triples \cite{ConnesLandi,ConnesDubois,Himalia,ChakPal,DabSit}, and more generally
the theory of pseudodifferential operators \cite{DiSj,Shubin},
the construction of star product \cite{Kon,Selene}, integrable systems etc. For reviews
on these topics, see \cite{Book, Polaris, Madore, Landi, Landsman}.

It has been proposed for a long time \cite{Snyder,Yang}
that the noncommutative spacetime is a quantum
effect of gravity and that this could provide some hints for the
regularization of quantum field theory. Naturally, many types of
action have been proposed and here we choose the spectral action introduced by
Chamseddine and Connes \cite{AliAlain1,AliAlain2} in a proper noncommutative
geometry setting. Since a similar action was derived in \cite{Gayral,Himalia}
following a prescription by Connes and Lott \cite{ConnesL1}, it is interesting to quote
the differences. In \cite{AliAlain2}, the idea is to recover the usual action of the
standard model of particle physics from purely gravitational considerations; more
precisely to define the bosonic action by $\Tr(\phi(\D^2/\Lambda^2))$ while the fermionic
action is simply $\langle \psi,\;\D \psi\rangle$ where $\D$ is the Dirac operator,
$\Lambda$ is an energy scale cut-off and $\phi$ is a smooth positive function. So
Chamseddine and Connes recovered the Einstein plus Yang--Mills and Weyl actions including
of course the spin 1 bosons, but also the part induced by the Higgs fields of spin 0.
The action functional computed in \cite{Gayral,Himalia} is defined by
$\Tr^+(F^2\D^{-d})$ where $F$ is the field strength curvature of a vector potential
and $\Tr^+$ is a Dixmier trace which pins down the leading term of logarithmic divergence
in the usual trace of $F^2$.

Here we choose for the manifold the Moyal plane $\R^{2m}$ with flat curvature,
technically viewed as an algebra $(\SS(\R^{2m}),\mop)$ for the Moyal product $\mop$ ($\SS$ is
the Schwartz space) and as Dirac operator, the usual one
$-i(\pa_{\mu}+L^\Th(\omega_{\mu})) \otimes
\gamma^{\mu}$ but where the connection $\omega$ acts by Moyal left multiplication
$L^\Th(\omega)$ on the usual Hilbert space of $L^2$-sections of the trivial spinor bundle
spinors $\H=L^2(\R^{2m})\otimes\C^{2^m}$. In \cite{Himalia}, this algebra and one of its
unitizations is proved to be a real spectral triple
of spectral dimension $2m$. So this case completely
fits the requirements and the computation of
$\Tr(\phi(\D/\Lambda))$ is possible using a standard heat
kernel technique. The not so surprisingly result
is that one recovers the usual commutative action
where all commutative products have been replaced by the Moyal ones.

Despite the fact that this
computation is made here in Euclidean signature and with no real gravity, the main
interest is that all the algebraic and analytical difficulties are overcome and that it
is the first example of spectral action for a not almost commutative spectral triple.
Since the Dirac operator
has a noncompact resolvent, a spatial regularization by the multiplication of a function
$f$ in the algebra is introduced to get the tracability of $L^\Th(f)e^{-t \triangle^\Th}$
where $\triangle^\Th$ is the generalized Moyal Laplacian. Of course, the choice of a regularization
is arbitrary and one could prefer for instance a soliton one \cite{LLS}, when one wants to
avoid, before the limit, the renormalization problems set in by UV{\tt /}$\,$IR mixing.

After some reminders on the role of the Moyalology for spectral triples in
section 2, we establish the main result, and then, give the important technical
details on the heat kernel computation in section 3. Section 4 is just an
application to the case of Moyal planes which ends with few remarks
on the difficulties with non flat cases.

\section{Moyal spectral triples and spectral action}
\subsection{Moyal analysis}

In this section, the very basic tools on Moyal analysis are recalled and we refer to
\cite{Himalia} for a review. The Moyal product comes from the
phase space formulation of  flat quantum mechanics, that is a
deformation of the associative algebra structure
of a suitable family of functions on $\R^{2m}$ with pointwise
product in the direction of the flat Poisson
bracket $\{.,.\}_P$. More precisely, if we denote by $W$ the
Weyl map which assigns Schwartz functions on $\R^{2m}$ say,
to bounded operators on $L^2(\R^m)$, the Moyal product $\mop$
is constructed in order to obtain a $*$-algebra homomorphism
$$
W:\SS(\R^{2m})\rightarrow\L\left(L^2(\R^m)\right),\;\;
W(f\mop g)=W(f) \; W(g).
$$
This leads us to
\begin{equation}
(f \mop g)(x) := (\pi\th)^{-2m} \iint_{\R^{2m}\times\R^{2m}} f(y) \; g(z)\,
e^{\frac{2i}{\th}(x-y)\,\.\,S(x-z)}\;  d^{2m}y \;d^{2m}z,
\label{eq:moyal-prod}
\end{equation}
where $S:= \begin{pmatrix} 0 & 1_m \\ -1_m & 0
\end{pmatrix}$ comes from the canonical symplectic structure of
$\R^{2m}=T^*\R^m$ and $\th\in\R^*_+$ is the deformation parameter.

Actually, one can define Moyal products on $\SS(\R^n)$, $n$ even
or odd, independently of any symplectic structure. In those cases,
it is any real skewsymmetric matrix $\Th$ which defines the deformation
directions. Generic Moyal products are then defined by
\begin{eqnarray}
\label{mogeneral}
(f \Mop g)(x):=(2\pi)^{-n}\int_{\R^{n}\times\R^{n}}  e^{i\xi(x-y)}
f(x-\textstyle{\frac{1}{2}}\Theta \xi) \;g(y)\;d^ny\; d^n\xi.
\end{eqnarray}

To fix notations and avoid to refer to the odd or even case,
$n$ will be an integer equals to the plane dimension
and $m$ will be the integer part of $\frac{n}{2}$.

This formula shows that the theory
of pseudodifferential operators on $\R^n$ \cite{Hor,Shubin}
is suitable for the analysis of left
and right Moyal multiplication operators $L^\Th(f)$ and
$R^\Th(f)$ defined by $L^\Th(f)\psi:=f \Mop
\psi$ and $R^\Th(f)\psi:= \psi \Mop f$. In particular their symbols are
\begin{equation}
\label{symbol}
\sigma\left[L^\Th(f)\right](\xi,x)=f(x-\thalf\Th\xi),\;\;\;
\sigma\left[R^\Th(f)\right](\xi,x)=f(x+\thalf\Th\xi).
\end{equation}
On coordinate functions $x^\mu,\;\mu=1,\cdots,n$,
generic Moyal products formally define
a generalized Heisenberg Lie algebra
structure:
$$
\left[x^\mu,x^\nu\right]_{\Mop}=i\Th^{\mu\nu}1.
$$
This equality can have a precise analytical meaning if we work
with Moyal products on some tempered distribution spaces
\cite{Phobos, Deimos}, and is obvious from the asymptotic expansion
of Moyal products:
\begin{equation}
\label{Mopas}
f\Mop g\sim\sum_{\alpha\in\N^n} \textstyle{\left(\frac{i}{2}\right)^{|\alpha|}
\frac{1}{\alpha !} } \; \frac{\pa f}{\pa x^\alpha} \; \frac{\pa g}{\pa (\Th
x)^\alpha}.
\end{equation}
This expansion can be heuristically derived from (\ref{mogeneral})
by a Taylor expansion of $\sigma\left[L^\Th(f)\right](\xi,x)$ "near" $x$,
for a  more rigorous approach see \cite{EGV}.
Moyal products satisfy a few useful algebraic equalities (see \cite{Himalia}
for a review); in particular the Leibniz rule is satisfied, the integral is a
faithful trace and the complex conjugation is an involution:
\begin{eqnarray}
\label{leibniz}
\pa^\mu(f\Mop g)&=&\pa^\mu f\Mop g+f\Mop\pa^\mu g,\\
\label{trace}
\int\; (f\Mop g)(x)\; d^{n}x&=&\int \; f(x) \; g(x)\;d^{n}x,\\
(f\Mop g)^*&=&g^*\Mop f^*.
\end{eqnarray}
These properties allow to prove that $\B_\Th:=\left(\SS(\R^n),\Mop\right)$
is an associative and involutive Fr\'echet algebra with a jointly
continuous product.

For $\Th$ symplectic, hence $n=2m$, it is proved in
\cite{Phobos, Deimos} that $\left(L^2(\R^{2m}),\mop\right)$ is an
associative Banach algebra, and we have shown in \cite{Himalia}
that $\left(\D_{L^2}(\R^{2m}),\mop\right)$ is also an $*$-algebra
with a jointly continuous product, $\D_{L^2}(\R^{2m})$ being the space
of smooth functions, having all their derivatives in $L^2(\R^{2m})$
endowed with the Fr\'echet topology of $L^2$-convergence for all
derivatives. For the nonunital spectral triple point of view (see below),
one needs also to choose a unitization for these algebras. In \cite{Himalia}
is studied the unital $*$-algebra $\left(\Oh_0(\R^{2m}),\mop\right)$,
where $\Oh_0(\R^{2m})$ consists in smooth bounded functions
with bounded derivatives, with the topology given by the family of semi-norms
$\{p_\alpha\}_{\alpha\in\N^{2N}}$, $p_\alpha(f):=\|\pa^\alpha f\|_\infty$.

For generic Moyal products, $\left(\SS(\R^n),\Mop\right)$,
$\left(\D_{L^2}(\R^n),\Mop\right)$, $\left(\Oh_0(\R^n),\Mop\right)$
are also Fr\'echet algebras. Actually this statement comes from
the algebra structure of $\SS(\R^n),\D_{L^2}(\R^n),\Oh_0(\R^n)$
with pointwise product as well as with Moyal one, and because any
Moyal product splits into a symplectic Moyal product and a pointwise one
denoted by a point \cite[Proposition 2.7 and Corollary 2.8]{RieffelBook}: Namely,
$f\Mop g(x)=f(x)\,.\,g(x)$ when $\Th=0$, so if the matrix $\Th$ is decomposed
as the direct sum of a symplectic one $\th$ of dimension $2m$
and the zero matrix of dimension $n-2m$, then
$\left(\SS(\R^n),\Mop\right) \cong \left(\SS(\R^{2m}),\mop\right)
\widehat{\otimes} \left(\SS(\R^{n-2m}),.\right)$. Remark in
particular that
$\SS(\R^n)$ and $\Oh_0(\R^n)$ are algebras
for pointwise product while for $\D_{L^2}(\R^n)$ this is a consequence
of the inclusion $\D_{L^2}(\R^n)\subset\Oh_0(\R^n)$ \cite{Schwartz}.

A {\it spectral triple}  $(\A,\tilde{\A},\H,\D,J,\chi)$ (a noncommutative
generalization of a Riemannian spin manifold) consists of an
algebra $\A$, a suitable one of  its unitizations $\tilde{\A}\supset\A$
(for the analogue of the noncompact case) both faithfully
represented by bounded operators on a separable Hilbert
space $\H$ (the representation is denoted by $\pi$), together
with an unbounded selfadjoint operator $\D$ such that $\pi(a)(\D+\lambda)^{-1}$
is a compact operator for all $a\in\A$ and $\lambda$ in the
resolvent of $\D$ and such that the commutators $[\D,\pi(a)]$ for all
$a\in\tilde{\A}$ extend to bounded operators. $J$ and $\chi$ are respectively
antiunitary and unitary operators with commutation relations
depending on the dimension of the triple. These data
must moreover fulfill a set of axioms (see \cite{ConnesReal, Himalia}).

It is shown in \cite{Himalia} that symplectic Moyal planes yield
nonunital spectral triples if we choose
$\A=\left(\D_{L^2}(\R^{2m}),\mop\right)$,
$\tilde{\A}=\left(\Oh_0(\R^{2m}),\mop\right)$, represented by the diagonal
left regular representation $\pi^\th(f):=L^\th(f)\otimes 1_{ 2^m}$ on the
Hilbert space of $L^2$-sections of the trivial spinor bundle
$\H=L^2(\R^{2m})\otimes\C^{2^N}$, and for $\D$, the flat Dirac operator
$\dslash:=-i\pa_\mu\otimes\gamma^\mu$ where $\gamma^\mu$ are
the Clifford matrices associated to $(\R^{2m},\eta)$ with $\eta$ the
standard Euclidean metric of $\R^{2m}$.

\subsection{Main result}

The {\it action functional} or
Connes--Lott action \cite{ConnesL1} associated with this spectral triple  gives the
noncommutative Yang--Mills action for symplectic Moyal products:
\begin{equation}
\label{NCYM}
YM(\alpha)=c\int F^{\mu\nu}\mop F_{\mu\nu}\;d^{2m}x,
\end{equation}
where $\alpha$ is a universal represented connection,
$$
\alpha\in\tilde{\pi}^\th\big(\{a_0\delta a_1\; : \;a_0,a_1\in\B_\th\}\big)=
\left\{\pi^\th(a_0)[\dslash,\pi^\th(a_1)]\;:\;a_0,a_1\in\B_\th\right\},
$$
and $F$ is its curvature:
$F^{\mu\nu}=\pa^\mu A^\nu -\pa^\nu A^\mu+[A^\mu,A^\nu]_{\mop}$ and
$A^\mu$ being defined by $\alpha=L^\th(A_\mu)\otimes\gamma^\mu$.
This result comes from the Junk computation \cite{Gayral} and the
following result \cite{Himalia}:

\begin{thm}
\label{pr:calcul}
For $f \in \SS(\R^{2m})$, the compact operator
$\pi^\th(f)\,(\Dslash^2 + \eps^2)^{-m}$ lies in
$\L^{(1,\infty)}(\H)$ and any of its Dixmier traces $Tr_\omega$
is independent of the positive number~$\eps$. More precisely we have,
\begin{equation}
\Tr_\omega \bigl( \pi^\th(f)\,(\Dslash^2 + \eps^2)^{-m} \bigr)
= {\textstyle{\frac{2^m\,\Omega_{2m}}{2m\,(2\pi)^{2m}}}} \int f(x) \,d^{2m}x,
\label{eq:cojoformula}
\end{equation}
where $\L^{(1,\infty)}(\H)$ is the ideal of compact operators whose
k-th singular values satisfy $\mu_k(T)=O(k^{-1})$ and $\Omega_{2m}$
is the hyper-area of the unit sphere in $\R^{2m}$.
\end{thm}

For generic Moyal products, $\left(\left(\D_{L^2}(\R^n),\Mop\right),
\left(\Oh_0(\R^n),\Mop\right),L^2(\R^n)\otimes\C^{2^m},
\dslash\right)$ yields also a
nonunital spectral triple, but the Connes--Lott action computation
is not obvious because the computation of (\ref{NCYM})
was basis dependent.

Let $\triangle^\Th$ be a {\it noncommutative generalized Laplacian} associated
with Moyal products
\begin{eqnarray}
\label{genelap}
\triangle^\Th &:=&-\Big(\eta^{\mu\nu}(\pa_\mu+L^\Th(\omega_\mu))
(\pa_\nu+L^\Th(\omega_\nu))+L^\Th(E)\Big)\otimes1_{2^m}, \\
\triangle^\Th &=:&\triangle_r^\Th \otimes 1_{2^m} \nonumber
\end{eqnarray}
acting on the Hilbert space $\H=L^2(\R^n)\otimes\C^{2^m}=:\H_r\otimes\C^{2^m}$, where
$\omega_\mu^*=-\omega_\mu$ and $E$ are in $\Oh_0(\R^n)$.
From now, let $\B_\Th:=\Big(\SS(\R^n),\Mop\Big)$ acting on $\H$ by the
diagonal left regular representation $\pi^\Th(.):=L^\Th(.)\otimes
1_{2^m}$.

For $f\in\B_\Th$, $L^\Th(f)e^{-t\,\triangle^\Th}$ will be called {\it spatially regularized
heat operator} associated with the generalized
Laplacian $\triangle^\Th$.

The following is the main result.
\begin{thm}
\label{HKNCP}
Let $\triangle^\th$ be as in (\ref{genelap}) and $f\in\B_\Th$. Then $\Tr
\left(\pi^\Th(f)\,e^{-t\,\triangle^\Th}\right)$ has an asymptotic expansion
\begin{equation}
\label{asympexp}
\Tr\left(\pi^\Th(f)\,e^{-t\,\triangle^\Th}\right)\sim_{t\rightarrow 0}\;
{\textstyle{2^m (\;\frac{1}{4\pi t})^{n/2}}}\sum_{l\in\N}
\;t^l\;\int_{\R^n}f(x)\;\tilde{a}_{2l}(x) \; d^nx,
\end{equation}
where
\begin{eqnarray*}
\tilde{a}_0(x)&=&1,\\
\tilde{a}_2(x)&=&E(x),\\
\tilde{a}_4(x)&=&\tfrac{1}{2}\; E\Mop E(x)+\tfrac{1}{6}\;
\eta^{\mu\nu} E_{;\mu\nu}(x)+\tfrac{1}{12}\;
\Omega^{\mu\nu}\Mop \Omega_{\mu\nu}(x),\\
\tilde{a}_6(x)&=&\tfrac{1}{6}\; E\Mop E\Mop E(x)
+\tfrac{1}{12}\;\eta^{\mu\nu}E_{;\mu}\Mop E_{;\nu}(x)
+\tfrac{1}{6}\;\eta^{\mu\nu} E\Mop E_{;\mu\nu}(x)\\
&&\hspace{-1.5cm}+\tfrac{1}{60}\;\eta^{\mu\nu}\eta^{\rho\sigma}
 E_{;\mu\nu\rho\sigma}(x)
+\tfrac{1}{12} \;E\Mop \Omega^{\mu\nu}\Mop \Omega_{\mu\nu}(x)
+\tfrac{1}{45}\;\eta^{\rho\sigma}{ \Omega^{\mu\nu}}_{;\rho}
\Mop \Omega_{\mu\nu ;\sigma}(x)\\
&&\hspace{-1.5cm}+\tfrac{1}{180}\;{\eta^{\rho\sigma} \Omega^{\mu\nu}}_{;\nu}
\Mop\Omega_{\mu\rho;\sigma}(x)
+\tfrac{1}{30}\;\eta^{\rho\sigma} \Omega^{\mu\nu}
\Mop \Omega_{\mu\nu;\rho\sigma}
-\tfrac{1}{30}\; \Omega^{\mu\nu}
\Mop \Omega_{\nu\rho}\Mop {\Omega^\rho}_\mu(x),
\end{eqnarray*}
where $g_{;\mu}:=\pa_{\mu}g+ [\omega_{\mu},g]_{\Mop}$ and
$\Omega_{\mu\nu}:=\pa_{\mu}\omega_{\nu}-\pa_{\nu}\omega_{\mu}+[\omega_{\mu},\omega_{\nu}]_{\Mop}$
is the curvature of the connection $\omega$.
\end{thm}

\section{Heat kernel expansion for Moyal generalized Laplacians}

We will first discuss the heat kernel expansion for Laplace type operators
associated with Moyal products, for NC planes as well as for NC tori.
This section generalizes Vassilevich's result \cite{Vass} in two directions: first the
Moyal products are defined by their integral form as opposed to differential or formal
Moyal products (\ref{Mopas}) and second they are taken over the whole plane $\R^n$ and not only
on NC tori. This noncompact situation generates some analytical difficulties.

We use the standard one-parameter semigroup theory of
$e^{-t\,A}$ where $A$ is a positive (unbounded) operator
and $t\in \R^+$.
Let $\H$ be a separable Hilbert space. We denote by $\B(\H)$ the set
of bounded operators on $\H$, by $\K(\H)$ the compact one's and by
$\L^p(\H)$ the $p$-th Schatten class.

If we assume that $A$ is a nonnegative
selfadjoint operator on $\H$, then $e^{-z\,A}$ is holomorphic for $\Re(z)>0$
and $\| e^{-z\,A} \| \leq1$ \cite[Example 1.25, p. 493]{Kato} \cite{Zagreb}.
With $\mathcal{R}_A(z):=(z-A)^{-1}$ denoting the resolvent of $A$, one can use the holomorphic functional
calculus:
\begin{eqnarray}
\label{Hcal}
e^{-t\,A}=\frac{1}{2i\pi}\int_{\Gamma}e^{-tz}\;\mathcal{R}_A(z) \; dz,
\end{eqnarray}
where $\Gamma$ is a positively oriented (possibly infinite) closed curve containing the spectrum
of $A$.

\begin{lem}
\label{zagreblike}
Let $B$ be a bounded operator and $A$ be  a nonnegative densely defined generator
of a holomorphic semigroup such that  $B\mathcal{R}_A(z)^l\in \L^1(\H)$ for some
$z\notin \spec(A)$.
Then for $t>0$, $Be^{-t\,A}$ is trace-class.
\end{lem}
\begin{proof}
For some $z_0\notin \spec(A)$, the semigroup property together with the first resolvent
equation and (\ref{Hcal}) gives:
$$
B\;e^{-t\,A}=B\; (e^{-\frac{t}{l}\,A})^l=B\;\mathcal{R}_A(z_0)^l\;\left(\frac{1}{2i\pi}
\int_\Gamma e^{-\frac{t}{l}z}(1+(z-z_0)\mathcal{R}_A(z))\;dz \right)^l.
$$
This concludes the proof because $\| \mathcal{R}_A(z) \| \leq \frac{M}{|z|}$ for
all $z$ with $\Re(z)>0$, thus
\begin{align}
\int_\Gamma e^{-\frac{t}{l}\,\Re(z)}\Big(1+|z-z_0|\;\|\mathcal{R}_A(z)\|\Big)\;|dz|\;<\infty.
\tag*{\qed}
\end{align}
\hideqed
\end{proof}
Since we are interested in the small $t$-asymptotic expansion of
$\Tr(Be^{-t\,A})$, recall the following
definition:

Let
$\{f_n\}_n$ be a sequence of functions such that
 $f_n(t) \neq 0$ for $t \neq 0$ and $f_{n+1}(t)=o(f_n(t))$ as $t\rightarrow 0$.
A function $f$ has the {\it asymptotic expansion} $f(t) \sim_{t\rightarrow 0} \sum_{n=0}
^{\infty}
a_n\;f_n(t)$,  when for each $k \in \N$, $f(t)= \sum_{n=0}^k a_n\;f_n(t) + O(f_{k+1}(t))$
as $t\rightarrow 0$.

\subsection{Heat kernel expansion for Moyal planes}

We will first show that $L^\Th(f)e^{-t\;\triangle_r^\Th}$ is
trace-class for $t\in\R^*_+$, then we will show that its trace has a small $t$-asymptotic
expansion:
\begin{equation}
\Tr\left( L^\Th(f)e^{-t\;\triangle_r^\Th}\right) \sim_{t\rightarrow 0}\;{(\textstyle{\frac{1}{4\pi
t})^{n/2}}}\sum_{l\in\N}
\;t^l\;\int_{\R^n}\;f(x)\tilde{a}_{2l}(x) \; d^nx,\nonumber
\end{equation}
where the local invariants $\tilde{a}_l$ are built from
the universal (represented) connection $\omega_\mu$,
the (nonlocal) endomorphism $E$ and their covariant derivative (in the adjoint representation)
$\pa_\mu+L^\Th(\omega_\mu)-R^\Th(\omega_\mu)$.

We will prove that $L^\Th(f)e^{-t\triangle_r^\Th}$ is trace-class by two approaches. The first uses
semigroup theory results while the second will be based on pseudodifferential operator
($\Psi$DO) techniques, which is more in the spirit of \cite{Himalia}.
\begin{thm}
\label{traceclass}
Let $f\in\SS(\R^n)$,  $\omega_\mu ,E\in\Oh_0(\R^n)$ with
$\omega_\mu^*=-\omega_\mu$ and $E=-g^*\Mop g$ for some
$g\in\Oh_0(\R^n)$. Then, for all $t>0$ the spatially regularized heat operator associated
with $\triangle^\Th$ defined in (\ref{genelap}) is trace-class.
\end{thm}

\begin{proof}[First proof of Theorem \ref{traceclass}]
Because $(L^\Th(g))^*=L^\Th(g^*)$, $\Im(g)=0$ implies  $L^\Th(g)$ is selfadjoint, so
$\triangle^\Th$ is positive. Thanks to Lemma \ref{zagreblike},
it is enough to prove that $L^\Th(f)R_{\triangle_r^\Th}(z)^l$ is trace-class for $l>\frac{n}{2}$.

Let us anticipate further notations to see that $\triangle^\Th$ is a squared covariant Dirac
operator:
\begin{eqnarray*}
\triangle^\Th &=&\dslash^2_\omega -B,\cr
\dslash_\omega &:=&-i\big(\pa_\mu+L^\Th(\omega_\mu)\big) \otimes \gamma^\mu,
\end{eqnarray*}
and $B:=L^\Th(E)\otimes1_{2^m}-L^\Th(\pa_\mu(\omega_\nu)-\omega_\mu\Mop\omega_\nu)
\otimes(\eta^{\mu\nu}1_{2^m}-\gamma^\nu\gamma^\mu)$ is bounded.

Assume first that $l=1$, $z=-1$, $B=0$. Using the notations $\pi^\Th(\omega):=L^\Th(\omega_\mu)
\otimes\gamma^\mu$, $\pi^\Th(\dslash(f)):=L^\Th(\pa_\mu f)\otimes\gamma^\mu$
and the fact that all $f\in\SS(\R^n)$ factorizes as $f=f_1\Mop f_2$,
for some $f_1,f_2\in\SS(\R^n)$ \cite[Proposition 2.7]{Himalia}, one gets:
\begin{eqnarray*}
\pi^\Th(f)\mathcal{R}_{\triangle^\Th}(-1)&=&-\pi^\Th(f)\frac{1}{\dslash-i}
\left(1-\pi^\Th(\omega)\frac{1}{\dslash_\omega-i}\right)\frac{1}{\dslash_\omega+i}\\
&=&-\pi^\Th(f_1)\frac{1}{\dslash-i}\;\pi^\Th(f_2)\left(1-\pi^\Th(\omega)\frac{1}{\dslash_\omega-i}\right)
\frac{1}{\dslash_\omega+i} \\
&&-\pi^\Th(f_1)\frac{1}{\dslash-i}\;\pi^\Th(\dslash(f_2))\frac{1}{\dslash-i}
\left(1-\pi^\Th(\omega)\frac{1}{\dslash_\omega-i}\right)
\frac{1}{\dslash_\omega+i} \\
&=&-\pi^\Th(f_1)\frac{1}{\dslash-i}\;\pi^\Th(f_2)\frac{1}{\dslash+i}
\left(1-\pi^\Th(\omega)\frac{1}{\dslash_\omega+i}\right)\\
&&+\pi^\Th(f_1)\frac{1}{\dslash-i}\;\pi^\Th(f_2)\Mop\omega)\frac{1}{\dslash-i}
\left(1-\pi^\Th(\omega)\frac{1}{\dslash_\omega-i}\right)\frac{1}{\dslash_\omega+i}\\
&&-\pi^\Th(f_1)\frac{1}{\dslash-i}\;\pi^\Th(\dslash(f_2)\Mop\omega)\frac{1}{\dslash-i}
\left(1-\pi^\Th(\omega)\frac{1}{\dslash_\omega-i}\right)\frac{1}{\dslash_\omega+i}\;.
\end{eqnarray*}
By \cite[Lemmata 4.5 and 4.14]{Himalia},
$\pi^\Th(g)\mathcal{R}_\dslash(i) \; \pi^\Th(h)\mathcal{R}_\dslash(i) \in \L^p(\H)$
whenever
$g,h\in\SS(\R^n)$ and $p>\frac{n}{2}$. The boundness of $\pi^\Th(\omega)$ and
$\mathcal{R}_{\dslash_\omega}(z)$ then yields $\pi^\Th(f)\mathcal{R}_{\triangle^\Th}(-1)\in \L^p(\H)$
for the same $p$.
For $l\geq 1$, repeat this algorithm using $(A+C)^{-1}=A^{-1}(1-CA^{-1}(\cdots(1-C(A+C)^{-1})\cdots))$
up to order $l$.

The case with non-zero $B$ is obtained using the same trick:
$$
\pi^\Th(f)\,\mathcal{R}_{\triangle^\Th}(-1)=\pi^\Th(f)\,\mathcal{R}_{\dslash_\omega^2}(-1)\left(
1-B\,\mathcal{R}_{\triangle^\Th}(-1)\right).
$$
The first resolvent equation implies the same result for any $z$ instead of $-1$ in
the resolvent set of $\spec(\triangle^\Th)$.
\end{proof}

The second proof, which is based on a functional calculus for $\Psi$DO \cite{DiSj},
needs the following definition of $\Psi$DO classes relevant for Moyal
analysis (Shubin \cite{Shubin} or GLS \cite{GLS} symbol classes).

\begin{defn}
Let $S^{\rho,\lambda}$ be the Shubin or GLS symbol class
\begin{eqnarray*}
S^{\rho,\lambda}&:=&\left\{\sigma\in\CC(\R^{2n}):\forall \alpha,\beta\in\N^n,
\exists C_{\alpha\beta} \in \R^+ \right.\\
&&\hspace{3cm}\left.\left|\pa_x^\alpha\pa_\xi^\beta\sigma(\xi,x)\right|\leq C_{\alpha,\beta}
(1+|x|^2)^{(\rho-|\alpha|)/2}(1+|\xi|^2)^{(\lambda-|\beta|)/2}\right\},
\end{eqnarray*}
and let $\Psi^{\rho,\lambda}:=\left\{A\in\Psi DO:\sigma[A]\in S^{\rho,\lambda}\right\}$
be the associated $\Psi$DO class.
\end{defn}
Actually, $S^{\rho,\lambda}$ fits into the general H\"ormander symbol classes (see \cite[Chapter XVIII]{Hor}) $S(m,g)$
with order function $m(\xi,x)=(1+|x|^2)^{\rho/2}\;(1+|\xi|^2)^{\lambda/2}$ and slowly
varying metric $g_{\xi,x}=(1+|\xi|^2)^{-1}|d\xi|^2+(1+|x|^2)^{-1}|dx|^2$.

\begin{proof}[Second proof of Theorem \ref{traceclass}]
First, equation (\ref{symbol}) and the product formula for $\Psi$DOs allows us to compute
the symbol of $\triangle_r^\th$:
$$
\sigma[\triangle_r^\th](\xi,x)=\eta^{\mu\nu}\Big(\xi_\mu\xi_\nu-2i\omega_\mu(x-\thalf\Th\xi)\xi_\nu
-i\pa_\mu\omega_\nu(x-\thalf\Th\xi) -\omega_\nu\Mop\omega_\nu(x-\thalf\Th\xi)\Big)
+E(x-\thalf\Th\xi),
$$
and because $\omega_\mu, E\in\Oh_0(\R^n)$, $\triangle_r^\th$ lies in $\Psi^{0,2}$.

Let $\{f_N\}_{N \in \N}$ be the family of smooth compactly supported functions defined by
$f_N(x):=\chi_{_N}(x)\;e^{-x},$
where $0\leq \chi_{_N} \leq 1$, $\chi_{_N} \in\CC_c(\R)$ with $\chi_{_N}(x)=0$ for
$x\in]-\infty,-\epsilon]\cup[N,+\infty[$ for a fixed
$\epsilon >0$ and $\chi_{_N}(x)=1$ for $x\in[0,N-\epsilon]$. First, \cite[Theorem 8.7]{DiSj}
yields $f_N(t \triangle_r^\th)\in\Psi^{0,-\infty}$, and basic estimates
(see \cite[Section 2.4]{Himalia} for details) gives   $L^\Th(f)\in\Psi^{-\infty,0}$
for all $f\in\SS(\R^n)$. Then, by \cite[Lemma 18.4.3]{Hor} one obtains
$L^\Th(f)f_N(t \triangle_r^\Th)\in\Psi^{-\infty,-\infty}$ and its symbol is in $\SS(\R^{2n})$.
Therefore,
\begin{eqnarray*}
C&:=&\sum_{|\alpha|+|\beta|\leq 2n+1}\|\pa_x^\alpha\pa_\xi^\beta\sigma
\left[L^\Th(f)f_N(t\triangle_r^\Th)\right]\|_1\\
&\leq&\sum_{|\alpha|+|\beta|\leq 2n+1}C_{\alpha,\beta}\int
(1+|x|^2)^{(-l-|\alpha|)/2}(1+|\xi|^2)^{(-k-|\beta|)/2}\; d^nx \;d^n\xi,
\end{eqnarray*}
for some $C_{\alpha,\beta}<\infty$ and all $l,k\in\N$, hence $C<\infty$.
Finally, \cite[Theorem 9.4]{DiSj} shows
$L^\Th(f)f_N(t \triangle_r^\Th)$ is trace-class for all $N\in\N$.
Looking at the estimates in the proof of \cite[Theorem 8.7]{DiSj}, one can
find constants $C_{\alpha,\beta}$ independent of $N$
(because $e^{-x}$ is rapidly decreasing when $x\rightarrow+\infty$, the right
support of $f_N$ plays no role), therefore one obtains that $L^\Th(f)f_N(t\triangle_r^\Th)$ is
trace-class uniformly in $N$.

To finish the proof, it remains to show that $\slim L^\Th(f)f_N(t\triangle_r^\Th)=L^\Th(f) e^{-t
\triangle_r^\Th}$, because \cite[Proposition 2]{DeSi} will ensure that $L^\Th(f)
e^{-t\triangle_r^\Th}$ is trace-class for all $t>0$.

Let  $\phi\in\H$ and $E_\lambda$ be the spectral family of $\triangle_r^\Th$, then
\begin{eqnarray*}
\|(\chi_{_N}(\triangle_r^\Th)-1)\phi\|_2^2&=&\langle\phi|(\chi_{_N}(\triangle_r^\Th)-1)^2\phi\rangle
=\int_{\spec(\triangle_r^\Th)}(\chi_{_N}(\lambda)-1)^2 \; d\langle\phi|E_\lambda\phi\rangle\\
&\leq&\int_{\spec(\triangle_r^\th)} \; d\langle\phi|E_\lambda\phi\rangle
=\langle\phi|\phi\rangle.
\end{eqnarray*}
Hence by dominated convergence and with $\phi=e^{-t\,\triangle_r^\Th}\psi$
\begin{eqnarray*}
\lim_{N\rightarrow\infty}\|(\chi_{_N}(\triangle_r^\Th)-1)\;e^{-t\;\triangle_r^\Th}\psi\|_2^2
&=&\lim_{N\rightarrow\infty}\int_{\spec(\triangle_r^\Th)}(\chi_{_N}(\lambda)-1)^2 \;
d\langle e^{-t\triangle_r^\Th}\psi, \;E_\lambda e^{-t\,\triangle^\Th}\psi\rangle\\
&=&\int_{\spec(\triangle_r^\th)}\lim_{N\rightarrow\infty}(\chi_{_N}(\lambda)-1)^2 \;
d\langle e^{-t\triangle_r^\Th}\psi,\;E_\lambda e^{-t\,\triangle_r^\Th}\psi\rangle=0,
\end{eqnarray*}
where the last equality comes from $\spec(\triangle_r^\Th)\subset\R^+$.
\end{proof}

We now come to the computation of the exponential of $\triangle^\Th$
following a Vassilevich's idea \cite{Vass}.
\begin{proof}[Proof of Theorem~\ref{HKNCP}] Let
\begin{eqnarray*}
X&:=&2L^\Th(\omega_\mu)\pa^\mu+L^\Th(\pa_\mu\omega^\mu)+
L^\Th(\omega_\mu\Mop\omega^\mu)+L^\Th(E)\cr
Y&:=&-\pa_\mu\pa^\mu,
\end{eqnarray*}
so $\triangle_r^\Th=Y-X$ and the Baker-Campbell-Hausdorff (BCH)
formula
\begin{equation*}
e^T\;e^S=e^{T\,+\,S\,+\frac{1}{2} \,[T,\,S]\,+\,\frac{1}{12}\,[T,\,[T,\,S]]\,
+\,\frac{1}{12}\,[S,\,[S,\,T]]\,-\,\frac{1}{48}\,[T,\,[S,\,[T,\,S]]]\,+\,\cdots}\;,
\end{equation*}
allows to write
\begin{equation*}
e^{-t\,\triangle_r^\Th}=e^{t\,X \, + \, \frac{1}{2} t^2\,[X,\,Y] \, + \, \frac{1}{12}t^3\,[X,\,[X,\,Y]] \,
- \, \frac{1}{6}t^3\,[Y,\,[X,\,Y]] \, - \, \frac{1}{48}t^4[X,\,[Y,\,[X,\,Y]]] \,
+ \, \frac{1}{48}t^4\,[Y,\,[Y,\,[X,\,Y]]] \, + \, \cdots}\; \;e^{-t \, Y}.
\end{equation*}
In order to obtain a power expansion when $t$ goes to zero,
the strategy is to expand the first exponential, to compute the commutators, to reorganize
the sequence and finally to write down the explicit symbol of those $\Psi$DO's. The trace
will be simply taken by integrating them with respect to $(\xi,x)\in\R^{2n}$.
Actually, the reorganization in homogeneous terms
in powers of $t$ is slightly more elaborate than a simple exponential expansion.
All the operators coming from this expansion are of the type
$L^\Th(g) \; \pa^\alpha,\;\alpha\in\N^n$, for some $g\in\B_\Th$.
Some terms will give no contributions to the trace since
$\int \xi_{1}^{\alpha_1}\cdots \xi_{n}^{\alpha_n}\;e^{-t|\xi|^2} \; d^n\xi=\Pi_i^n \;
\frac{1}{2}(1+(-1)^{{\alpha_i}})\;\Gamma(\frac{n+1}{2})\;t^{-(\alpha_i+1)/2}$ is zero when
at least one of the
$\alpha_i$ is odd, and when they are all even, $\vert\alpha\vert=\sum_i^n\alpha_i=2l$ is
even and we get
$$
\int_{\R^n} \xi_{\mu_1}\cdots \xi_{\mu_{2l}} \;e^{-t|\xi|^2}\; d^n\xi= {\textstyle{
\left(\frac{\pi}{t}\right)^{n/2}}} \; (2t)^{-l} \; \sum_{\sigma\in S_{2l}}
{\textstyle{\frac{1}{2^l l!} }}\; \delta_{\sigma(\mu_1)\sigma(\mu_2)}\cdots
\delta_{\sigma(\mu_{2l-1})\sigma(\mu_{2l})},
$$
where $\sigma$ runs over the permutation
group $S_{2l}$ of $2l$ elements. So, in the reorganization of
the  power series, we have to keep in mind that
$t^l \; L^\Th(g) \; \pa^\alpha$ is effectively a term of order
$t^{l-\frac{|\alpha|}{2}}$ (independently of the ${\textstyle{
\left(\frac{\pi}{t}\right)^{n/2}}}$ term.)
Moreover, to obtain the asymptotic expansion up to order $l$
say, we have to use the BCH formula up to order $2l-1$.
The order of the BCH formula is defined as the number of commutators
in the expansion. The term with higher degree derivatives
coming from the BCH formula at order $l$ is
$$
[t\,\pa^2,\;[t\,\pa^2,\;\cdots,\;[t\,\pa^2, \; t \,L^\Th(g)\pa ]\cdots]\;]\propto t^{l+1} \; L^\Th(h)
\;\pa^{l+1},
$$
for some $h\in\B_\Th$, which yields a term of order $t^{\frac{l+1}{2}}$.

Let us show how it works up to order one. We have to use the BCH formula up to order
one also: $e^{-t \, \triangle_r^\Th}=e^{t\,X\,-\,t\,Y}=e^{t\,X\,+\,\frac{1}{2}
[tX,\,tY]\,+\,\cdots}\;e^{-t\,Y}$, and
\begin{eqnarray*}
[tX,tY]&=&t^2\;\left[\pa_\nu\pa^\nu,2L^\Th(\omega_\mu)\pa^\mu+
L^\Th(\pa_\mu\omega^\mu) +L^\Th(\omega_\mu\Mop\omega^\mu)
+L^\Th(E)\right]\\
&=&t^2\Big(2L^\Th(\pa_\nu\pa^\nu\omega_\mu)\pa^\mu
+4L^\Th(\pa_\nu\omega_\mu)\pa^\mu\pa^\nu
+L^\Th(\pa_\nu\pa^\nu\pa_\mu\omega^\mu)
+2L^\Th(\pa_\nu\pa_\mu\omega^\mu)\pa^\nu\\
&&\quad +L^\Th(\pa_\nu\pa^\nu(\omega_\mu\Mop\omega^\mu))
+2L^\Th(\pa_\nu(\omega_\mu\Mop\omega^\mu))\pa^\nu
+L^\Th(\pa_\nu\pa^\nu E)+2L^\Th(\pa_\nu E)\pa^\nu\Big)\\
&=&4t^2\;L^\Th(\pa_\nu\omega_\mu)\pa^\mu\pa^\nu+O(t^2),
\end{eqnarray*}
hence
\begin{eqnarray*}
L^\Th(f)\,e^{-t \,\triangle_r^\Th}&=&L^\Th(f)e^{t\left(2L^\Th(\omega_\mu)\pa^\mu
+L^\Th(\pa_\mu\omega^\mu)+
L^\Th(\omega_\mu\Mop\omega^\mu)+L^\Th(E)\right)
+2t^2L^\Th(\pa_\nu\omega_\mu)\pa^\mu\pa^\nu+\cdots}\;
e^{t\pa_\mu\pa^\mu}\\
&=&L^\Th(f)\Big(1+t\big(2L^\Th(\omega_\mu)\pa^\mu+L^\Th(\pa_\mu\omega^\mu)+
L^\Th(\omega_\mu\Mop\omega^\mu)+L^\Th(E) \big)\\
&& \quad +2t^2\big( L^\Th(\pa_\nu\omega_\mu)\pa^\mu\pa^\nu
+ L^\Th(\omega_\mu\Mop\omega_\nu)\pa^\mu\pa^\nu\big)
+O(t^2)\Big) \;  e^{t\,\pa_\mu\pa^\mu}.
\end{eqnarray*}
where the last $t^2$-term comes from $e^{t\,X}$. So, by (\ref{symbol})
\begin{eqnarray*}
&&\hspace{-2.3cm} \sigma\left[L^\Th(f)\,e^{-t \,\triangle_r^\Th}\right](\xi,x) \\
&=&\Big(f(x-\thalf\Th\xi)+t\big(2f\Mop\omega_\mu(x-\thalf\Th\xi)(-i\xi)^\mu\\
&&\quad +\;f\Mop\pa_\mu\omega^\mu(x-\thalf\Th\xi)
+f\Mop\omega_\mu\Mop\omega^\mu(x-\thalf\Th\xi)
+f\Mop E(x-\thalf\Th\xi) \big)\\
&&\quad +\;2t^2\big( f\Mop\pa_\nu\omega_\mu(x-\thalf\Th\xi)(-i\xi)^\mu(-i\xi)^\nu\\
&&\quad+ \;f\Mop\omega_\mu\Mop\omega_\nu(x-\thalf\Th\xi)(-i\xi)^\mu(-i\xi)^\nu\big)
+O(t^2)\Big) \;e^{-t\,\xi_\mu\xi^\mu}.
\end{eqnarray*}

Finally, it remains to integrate $\sigma\left[L^\Th(f)\,e^{-t \,\triangle_r^\Th}\right](\xi,x)$.
By the translation $x\rightarrow x+\thalf\Th\xi$, one obtains:
\begin{eqnarray*}
&\hspace{-13cm}\Tr \left(L^\Th(f)\,e^{-t\,\triangle_r^\Th}\right)\cr
&\hspace{-0.4cm}=(2\pi)^{-n}\;{\displaystyle \iint}
\Big(f(x)+t\big(2f\Mop\omega_\mu(x)(-i\xi)^\mu +f\Mop\pa_\mu\omega^\mu(x)
+f\Mop\omega_\mu\Mop\omega^\mu(x)+f\Mop E(x) \big) \cr
&+\;2t^2\big( f\Mop\pa_\nu\omega_\mu(x)(-i\xi)^\mu(-i\xi)^\nu
+ f\Mop\omega_\mu\Mop\omega_\nu(x)(-i\xi)^\mu(-i\xi)^\nu\big)
\Big) \; e^{-t\,\xi_\mu\xi^\mu} \; d^nx \; d^n\xi \cr
&\hspace{-11.6cm} +\;O(t^{-n/2+2}) \cr
&\hspace{-1cm}=(4\pi t)^{-\frac{n}{2}}\;{\displaystyle \int}f(x)\;
\Big(1+t\big(\pa_\mu\omega^\mu(x)+
\omega_\mu\Mop\omega^\mu(x)+E(x) -\pa_\mu\omega^\mu(x)
-\omega_\mu\Mop\omega^\mu(x)\big)\Big)\;d^nx \cr
&\hspace{-11.8cm} +\;O(t^{{-\frac{n}{2}}+2}) \cr
&\hspace{-7.6cm}=(4\pi t)^{-\frac{n}{2}}\;{\displaystyle \int}f(x)\;
\big(1+tE(x)\big)\;d^nx\;+\;O(t^{{-\frac{n}{2}}+2}).
\end{eqnarray*}
The higher order terms can be obtained by similar computations,
that is to say, one generically get
$$
L^\Th(f)\,e^{-t\,\triangle_r^\Th}\sim_{t\rightarrow 0}
L^\Th(f)\Big(\sum_{l\in\N}\;t^l\;\sum_{\alpha\in\N^n,|\alpha|\leq l}
L^\Th(g_{\alpha,l})\;t^{|\alpha|/2}\;\pa^\alpha\Big) \; e^{t\,\pa_\mu\pa^\mu},
$$
for some $g_{\alpha,l}\in\B_\Th$, and where we have corrected
the powerseries in $t$ by the order of derivatives, with respect
to the previous discussion.
Here $\sim$  means asymptotic expansion with respect to the
trace-norm topology:
$$
\|L^\Th(f)e^{-t\triangle_r^\Th}-L^\Th(f)\Big(\sum_{l\leq
N}\;t^l\;\sum_{\alpha\in\N^n,|\alpha|\leq l}
L^\Th(g_{\alpha,l})\;t^{|\alpha|/2}\;\pa^\alpha\Big) \; e^{t\,\pa_\mu\pa^\mu}\|_1 =O(t^{N+1}),
$$
convergence of the sequence being warranted by Theorem \ref{traceclass}.

This concludes the proof of Theorem \ref{HKNCP} since in (\ref{asympexp}) we get the
coefficient $\Tr(1_{\C^{2^m}})$.
\end{proof}
\begin{rem}
This systematic computation also yields that the other coefficients
$\tilde{a}_{2l},\;l>3$ have the same canonical form, that is Moyal
products replace pointwise ones everywhere.
\end{rem}

\subsection{Heat kernel expansion for NC tori in Moyal (re)presentation}

Let $\A_\Th$ be the smooth algebra of Schwartz
(rapidly decreasing) linear combination of the plane
waves $\{e^{ik.x}\}_{k\in\Z^n}$ endowed with Moyal
product:
$$
\A_\Th=\left(\left\{\sum_{k\in\Z^n}c_k\; e^{ik.x}
\;:\;(c_k)\in\SS(\Z^n)\right\},\Mop\right).
$$
$\A_\Th$ closes to an algebra and represents the NC
$n$-tori:
\begin{equation}
e^{ik.x}\Mop e^{iq.x}=e^{-ik.\Th q}\;e^{iq.x}\Mop e^{ik.x},
\label{rccNCT}
\end{equation}
this canonical commutation relation of the NC $n$-tori
coming from the straightforward computation
(here Fourier modes are viewed as tempered distributions):
$$
e^{ik.x}\Mop e^{iq.x}=(2\pi)^{-n}\iint_{\R^n \times \R^n}\;
e^{i\xi.(x-y)}\;e^{ik.(x-\frac{1}{2} \Th \xi)}\;e^{iq.y}\;d^n\xi \;d^ny
=e^{-i\frac{1}{2} k.\Th q} \;e^{i(k+q).x}.
$$
One can build a unital spectral triple associated to this algebra
\cite{Gravity, Polaris}, with $\H=L^2(\T^n)\otimes\C^{2^m}$
 the squared integrable sections of the trivial spinor bundle over
 $\T^n$, and $\D=\dslash$ the flat Dirac operator. $\A_\Th$ is again
represented on bounded operators by the left regular representation
$\pi^\Th(a)\psi=L^\Th(a)\otimes 1_{2^m}\; \psi=a\Mop\psi$,
for $a\in\A_\Th, \psi\in\H$.
Actually, this construction is equivalent to the GNS representation
associated to the state given by the canonical trace $\tau$ of $\A_\Th$:
when $a(x)=\sum_{k\in\Z^n}c_k\;e^{ik.x}\in\A_\Th$,
\begin{equation*}
\tau(a):=c_0=\int_{\T^n} \;a(x)\;d^nx.
\end{equation*}

Let again $\triangle^\Th$ be the noncommutative generalized Laplacian
defined in (\ref{genelap}) acting now on $\H:=L^2(\T^n)\otimes\C^{2^m}$, where
$\omega_\mu^*=-\omega_\mu$ and $E \leq 0$ are in $\A_\Th$.

We will first show that in the NC-tori cases, $e^{-t\,\triangle^\Th}$ is
trace-class for
$t\in\R^*_+$ is a direct consequence of the compactness of
$\mathcal{R}_\triangle^\Th(z):=(\triangle^\Th-z)^{-1}$.
Then, thanks to the previous section, it will be straightforward
to show that its trace has a small-$t$ asymptotic
expansion (\ref{asympexp})
where the local invariants $\tilde{a}_l$ are the same as in the
Moyal plane case, but with $f=1$.
\begin{thm}
\label{tc}
Let $\triangle^\Th$ be as in (\ref{genelap}), then
$e^{-t \, \triangle^\Th}$ is trace-class for all $t\in\R^*_+$.
\end{thm}
\begin{proof}
The proof is simpler than for the Moyal plane. Clearly,
$\mathcal{R}_\dslash(z)\in\L^p(\H)$, $p>n$, and
so $\mathcal{R}_{\triangle^\Th}(z)\in\L^{p/2}(\H)$
(use the same trick as in the proof of Theorem
\ref{traceclass}). Then Theorem \ref{zagreblike} yields
$e^{-t \triangle^\Th}=(e^{-t\frac{2}{p}\triangle^\Th})^{\frac{p}{2}}
\in\L^1(\H)$.
\end{proof}

For the computation of the small-$t$ expansion, because all algebraic
properties (mainly Leibniz rule) used in the previous section work as
well as for the tori, we also obtain
$$
e^{-t\;\triangle^\Th}\sim_{t\rightarrow 0}\;
\sum_{l\in\N}\;t^l\;\Big(\sum_{\alpha\in\N^n,|\alpha|\leq l}
L^\Th(g_{\alpha,l})\;t^{|\alpha|/2}\;\pa^\alpha\Big) \; e^{t\,\pa_\mu\pa^\mu},
$$
where $g_{\alpha,l}\in\A_\Th$ are the same as for the Moyal plane case
except that now $f=1$. So the trace has the small-$t$ expansion:
\begin{eqnarray*}
\Tr\left(e^{-t\;\triangle^\Th}\right)&=&2^m(2\pi)^{-n} \;\int_{\T^n}d^nx
\int_{\R^n}\sigma
\left[e^{-t\;\triangle^\Th}\right](\xi,x)\;d^n\xi\\
&\sim&\hspace{-0.3cm}_{t\rightarrow 0}\;\;\tfrac{2^m}
{(2\pi)^{n}}\;\int_{\T^n}d^nx\int_{\R^n}
\sum_{l\in\N}\;t^l\;\Big(\sum_{\alpha\in\N^n,|\alpha|\leq l}
g_{\alpha,l}(x-\thalf\Th\xi)\;t^{|\alpha|/2}(-i\xi)^\alpha \Big)
\; e^{-t|\xi|^2}\;d^n\xi.
\end{eqnarray*}
Now, expanding $g_{\alpha,l}$ in Fourier modes and using
$$
\int_{\T^n}e^{ik(x-\frac{1}{2}\Th\xi)}\; d^nx=e^{-ik\frac{1}{2}\Th\xi}\;
\delta_{k,0}=\delta_{k,0},
$$
we directly obtain the following result:
\begin{thm}
\label{HKNCT}
Let $\triangle^\Th$ be as in (\ref{genelap}), then
\begin{equation*}
\Tr\left(e^{-t\,\triangle^\Th}\right)\sim_{t\rightarrow 0}\;
{\textstyle{2^m (\;\frac{1}{4\pi t})^{n/2}}}\sum_{l\in\N}
\;t^l\;\int_{\T^n}\;\tilde{a}_{2l}(x) \; d^nx,
\end{equation*}
where the $\tilde{a}_{2l}(x)$are given in Theorem \ref{HKNCP}.
\end{thm}

\section{The spectral action}
\subsection{Spectral action for nonunital spectral triples}

For a unital spectral triple $(\A,\H,\D)$, Chamseddine and
Connes \cite{AliAlain1, AliAlain2} proposed a definition of a physical action
which depends only on the spectrum of the covariant
$\D$-operator (the spectral action principle):
\begin{equation}
S(\D,A):=\Tr \left(\phi(\D_A^2/\Lambda^2)\right),
\end{equation}
where $\D_A$ is the covariant "Dirac" operator
$\D_A:=\D +A+\epsilon JAJ^{-1}$, $A$ is a universal
represented 1-form $A\in\tilde{\pi}(\Omega^1\A)$, $\tilde{\pi}$
being the lifted representation on the whole differential algebra
$\Omega^\bullet\A$ ($\tilde{\pi}(a_0\delta a_1\cdots\delta a_p)
:=\pi(a_0)[\D, \pi(a_1)]\cdots[\D, \pi(a_p)]$, $a_i\in\A, \;i=1,\cdots,p$),
$J$ is the real structure of the triple (the charge conjugation
for spinors in the commutative case), $\phi$ a suitable cut-off function,
$\Lambda$ a mass scale and $\epsilon\in\{+1,-1\}$ depending upon the
dimension. Any positive smooth function $\phi$
mimicking the step function $\chi_{[0,1]}$ was initially used in \cite{AliAlain1, AliAlain2} and
in \cite{Odysseus}, sufficient conditions on $\phi$ have been detailed.
Since in the unital case, $\D$ has compact resolvent
and likewise for the perturbed $\D_A$ by Theorem \ref{tc}, $\phi(\D_A^2/\Lambda^2)$
is trace-class as long as $\phi$ decreases fast enough; for instance $r^{n-1}\phi(r^2)\in L^1(\R^+)$
is a sufficient condition for a spectral triple with spectral
dimension equal to $n$.

Let us be more explicit about the covariant "Dirac" operator $\D_A$.
The starting point is the analogy between the invariance group of a
gauge theory on a Riemannian manifold $M$ coupled with
general relativity, $G=U\rtimes \Diff(M)$
and the group of automorphism of an algebra
$\A$ which splits into its inner and outer part $\Aut(\A)=\Int(\A)\rtimes\Out(\A)$,
with the following exact (group) sequence:
$$
1\rightarrow U\rightarrow G\rightarrow \Diff(M)\rightarrow 1,
$$
$$
1\rightarrow\Int(\A)\rightarrow\Aut(\A)\rightarrow\Out(\A)\rightarrow 1.
$$
In particular, if we choose $\A=\mathcal{C}^\infty(M, M_n(\C))
\cong\mathcal{C}^\infty(M)\otimes M_n(\C)$, $n>1$, the
two constructions coincide: $\Out(\A)=\Diff(M)$, $\Int(\A)=
\mathcal{C}^\infty(M,SU_n/\Z_n)$. The natural invariance group for
an action defined on a spectral triple must be the automorphism
group of the algebra. In order to retrieve a gauge theory with
spin matter when $\A$ is almost commutative that is
$\A=\mathcal{C}^\infty(M)\otimes A_F$ (where $A_F$ is a finite algebra such as
$\mathbb{H} \oplus\C \oplus M_3(\C)$ for the standard model of particle physics
\cite{AliAlain1,AliAlain2,CIKS}), we must represent $\Aut(\A)$ in the
fermionic Hilbert space $\H$. In particular, we have to lift $\Int(\A)$ to
the unitary group $\mathcal{U}(\H)$ of the bounded operators on $\H$:
$$
\mathcal{U}(\A)\ni u\mapsto \sigma(u)=\pi(u)J\pi(u)J^{-1}\in\mathcal{U}(\H).
$$
For NC tori, Moyal planes and some almost-commutative geometries,
this is the adjoint representation: $\pi(u)J\pi(u)J^{-1}\psi=
u\Mop \psi\Mop u^{*}$, $\psi\in\H$. Under this transformation,
$\D$ transforms as
\begin{equation}
\label{gaugetransform}
\D\rightarrow\sigma(u)\D\sigma(u)^{-1}=\D+\pi(u)[\D,\pi(u^*)]
+\epsilon J\pi(u)[\D,\pi(u^*)] J^{-1},
\end{equation}
where $\epsilon$ comes from commutation relations
$\D J=\epsilon J\D$, $\epsilon\in\{+1,-1\}$
(see \cite{ConnesReal, Polaris} for a table of signs)\footnote{The
sign $\epsilon$ in equation (\ref{gaugetransform})
is actually wrong in most of the literature,
however the computations linked with physics models are
unaffected because $\epsilon=1$ in the zero and four
dimensional cases.}. Hence $\D_A\rightarrow \D_{A'}$ with
$A'=\pi(u)A\pi(u^*)+ \pi(u)[\D,\pi(u^*)]$ transforms covariantly.

For almost commutative geometry $\mathcal{C}^\infty(M)\otimes A_F$,
in particular for the standard model, with
$\D=\Dslash\otimes 1_{\H_F}$ and the curved Dirac operator
$\Dslash=-i{e^\mu}_a \gamma^a(\pa_\mu+ \omega_\mu)$, $\omega$ being the spin connection
on $M$, $S(\D,A)$ is asymptotically computable by heat
kernel techniques. We may note that $\Dslash_A^2$ can be
written as a generalized Laplacian: $\Dslash_A^2=P$ with
$P=-\left(g^{\mu\nu}(\pa_\mu +\omega_\mu)(\pa_\nu +\omega_\nu)
+E\right)$ where $g^{\mu\nu}$ is the metric tensor, now $\omega_\mu$
is a connection containing spin and Yang--Mills part and $E$
is an endomorphism of the fiber bundle, on whose sections $P$ acts.
One can formally show \cite{AliAlain1,AliAlain2}, expanding $\phi$ in Taylor series,
that $S(\D,A)$ is linked to the Seeley--DeWitt coefficients $a_k(P,x)$
of the trace of the heat operator on a $n$-dimensional manifold
\begin{equation}
\Tr \left(e^{-tP}\right)\sim_{t\rightarrow 0} (4\pi)^{-n/2}\sum_{l\in\N}t^{(l-n)/2}\int_M a_l(P,x)
\; {\rm dvol}\/(x).
\end{equation}
where ${\rm dvol}\/(x)$ is the Riemannian volume form, by the relation
between zeta function and trace of the heat operator \cite{Gilkey}:
\begin{equation}
\label{zeta}
\zeta_P(s):=\Tr(P^{-s})={\textstyle{\frac{1}{\Gamma(s)}}}\int_0^\infty t^{s-1}
\Tr(e^{-tP})\;dt.
\end{equation}
On a manifold without boundary  $a_l(P,x)=0$,  $l$ odd, therefore
in the four dimensional case, this yields:
\begin{equation}
S(\Dslash\otimes 1_{A_F},A)= (4\pi)^{-2}\sum_{l=0}^{2}\Lambda^{4-2l}
\; \phi_{2l}\;\int_M a_{2l}(P,x) \;{\rm dvol}\/(x)\;+O(\Lambda^{-2}),
\end{equation}
where
\begin{equation}
\label{moments}
\phi_0=\int_0^\infty\phi(t)\;tdt,\;\;\phi_2=\int_0^\infty\phi(t)\;dt,\;\;
\phi_{2(2l+2)}=(-1)^l\phi^{(l)}(0), \;l\geq 0.
\end{equation}
A less formal, derivation of this relation
with precise constraints on $\phi$ can be found in \cite{Odysseus,
Nestetal}. For $M$ still four dimensional and
$A_F=\mathbb{H} \oplus\C \oplus M_3(\C)$, the spectral
action yields a unification of the Einstein plus Weyl gravity and
the standard model including the Higgs sector and its spontaneous
symmetry breaking (see \cite{AliAlain1,AliAlain2,CIKS}). There is no restriction for an arbitrary
dimension but the coefficients (\ref{moments}) will be slightly different, as we will
see below for the Moyal plane.
\begin{rem}
The relation (\ref{zeta}), links also the Dixmier trace with the heat
kernel expansion, and therefore the Connes--Lott action (\ref{NCYM}) with the
spectral action as explained in section \ref{comparaison}.
\end{rem}

For the nonunital case, since $\D$ has no longer a compact resolvent,
we invoke a {\it spatial regularization} $\rho$ to define the spectral action.
Like the energy regularization $\phi$, $\rho$ is a positive function rapidly decreasing,
in the almost commutative case, and
must be generically an element of the algebra $\A$.
\begin{defn}
For a nonunital spectral triple $(\A,\tilde{\A},\H,\D)$ of spectral
dimension $n$, the {\it spectral action} is
\begin{equation}
S(\D,A,\rho):=\Tr_\H\left(\pi(\rho)\;\phi(\D_A^2/\Lambda^2)\right),
\end{equation}
where as in the unital case, $\D_A=\D+A+\epsilon J AJ^{-1}$
($\epsilon\in\{+1,-1\}$ depending on $n$), and $A\in \Omega_{\D}^1(\tilde{\A})$ is now a represented
selfadjoint 1-form of the unitized algebra:
$A=\sum_{i\in I}\pi(b_0^i) \; [\D,\pi(b_1^i)]$, for $I$ a
finite set, $b_0^i,b_1^i\in\tilde{\A}$, $0\leq\rho\in\A$ and moreover
$0\leq\phi$, $\Lambda$ are as in the unital case.
\end{defn}

\begin{rem}
i) This definition gives more importance to the choice of the unitization.
The 1-form $A$ is now constructed from $\tilde{\A}$, and all
the symmetry considerations discussed previously occur now for the unitized
algebra. This is important because unitaries in the algebra are necessary to express gauge invariance:
$S(\D,A,\rho)$ is gauge invariant that is invariant under the lifted inner automorphism
implemented by the unitary operator $\pi(u)J\pi(u)J^{-1}$ and now the
regularization $\rho$ transforms also:
$$
\begin{cases}
A & \rightarrow \; uAu^*+u[\D,u^*], \\
\rho & \rightarrow \;u\rho u^* .
\end{cases}
$$
ii) The positivity of $\rho$ and $\phi$ is necessary in order to get a positive action.\newline
iii) Other regularizations are possible. For instance,
$\phi \big(\D_A^2\,\pi(\rho)^{-1} \big)$ where $f \in \SS$ is a strictly positive
function also give rise to trace-class operators for Moyal planes, but the
asymptotic expansion is still unmanageable.
\end{rem}
Let us show how it works for an almost commutative geometry associated
with a boundaryless noncompact smooth manifold $M$. In this case,
we still work with $\A=\mathcal{C}_c^\infty(M)\otimes A_F$. The operator $\rho \;e^{-t\,P}$,
$t\in\R^*_+$, is trace-class, for $\rho\in\mathcal{C}^\infty_c(M)$
viewed as a pointwise multiplication
operator and $P$ being a generalized Laplacian ($\Psi$DO operator of order two with metric tensor as
coefficient of the leading symbol). In this case the formula (\ref{zeta}) has an analogue:
\begin{equation}
\label{zeta2}
\zeta_{\rho,P}(s):=\Tr(\rho \;P^{-s})={\textstyle{\frac{1}{\Gamma(s)}}}\int_0^\infty t^{s-1}
\Tr(\rho \;e^{-t\,P})\;dt,
\end{equation}
one obtains, for $P=(\Dslash + A +\epsilon JAJ^{-1})^2$,
\begin{eqnarray*}
S(\Dslash\otimes 1_{A_F},A)= (4\pi)^{-n/2}\sum_{l=0}^m
\Lambda^{n-2l}\; \phi_{2l}\;\int_M a_{2l}(P,x)\;\rho(x) \; {\rm dvol}\/(x)
\;+\;O(\Lambda^{n-2(m +1)}),
\end{eqnarray*}
where $a_{2l}$ are still the Seeley--DeWitt coefficients which are now only locally
integrable while $a_{2l}(P,x)\rho(x)$ are globally integrable (see \cite{Vass2}). The coefficients
$\phi_{2l}$ have the form (\ref{moments}) in the four dimensional case, and their
values in any dimension $n$ is now computed for Moyal planes.

\subsection{The case of the Moyal plane}
Actually, the relation (\ref{zeta2}) is quite general, that is for any
bounded operator $S$ and any operator $T$
such that $S\;T^{-s}$ is trace-class, we have
\begin{equation}
\label{zeta3}
\zeta_{S,T}(s):=\Tr(S\;T^{-s})={\textstyle \frac{1}{\Gamma(s)}}\int_0^\infty
t^{s-1}\Tr\left(S \;e^{-t\,T}\right)\;dt.
\end{equation}
With this relation and the result of section 3, one can
derive the spectral action for Moyal planes. However in order to obtain more directly
the form of the coefficients $\phi_{2k}$ in any dimension, we will derive it by
Laplace transform techniques such as in \cite{Nestetal} (see \cite{Widder} for
details on Laplace transform).
We assume that the function $\phi$ has the following property:
\begin{equation}
\label{condition}
\phi\in\mathcal{C}^\infty(\R^+) \!\!\! \sepword {is the Laplace transform of} \!\!\!
\hat{\psi} \in \SS(\R^+):=\set{g\in\SS: g(x)=0, x \leq 0}
\end{equation}
 Thus, any function with this property has necessarily an analytic extension on
the right complex plane and is a Laplace transform.
Consequently, any $m$-differentiable function $\psi$ such that $\psi^{(m)}=\phi$ is the
Laplace transform of a function $\hat{\psi}$ and by differentiation, it satisfies
\begin{equation*}
 \phi(z)=\psi^{(m)}(z)=(-1)^m\int_0^\infty e^{-sz}\; s^m\;\hat{\psi}(s)\;ds,\;\Re z>0.
\end{equation*}
With $\triangle^\Th$ defined in (\ref{genelap}), using $\phi(\triangle_r^\Th)=(-1)^m\int_0^\infty
e^{-s\,\triangle_r^\Th}\,s^m\,\hat{\psi}(s) \;ds$ and the positivity of $\rho=g^*\Mop g$,
$g\in\B_\Th$, we get
$$
\Tr\left(L^\Th(\rho)\;\phi\left(\triangle_r^\Th/\Lambda^2\right)\right)=(-1)^m
\Tr\left(L^\Th(g)\int_0^\infty e^{-t\,\triangle_r^\Th/\Lambda^2}t^m\hat{\psi}(t) \; dt\;L^\Th(g^*)\right).
$$
Let $\{\Phi_p\}_{p\in\N}$ be any orthonormal basis of $\H_r$ and let
$0\leq B_t :=L^\Th(g)\,e^{-t\,\triangle_r^\Th/\Lambda^2}\,L^\Th(g^*)$, then
\begin{eqnarray*}
\Tr\left(L^\Th(\rho)\;\phi\left(\triangle_r^\Th/\Lambda^2\right)\right)&=&
\lim_{N\rightarrow \infty} \int_0^\infty\;
\sum_{p\leq N}\langle\Phi_p, \;B_t \;\Phi_p\rangle \;t^m\hat{\psi}(t)\;dt\\
&\leq&\lim_{N\rightarrow\infty}\int_0^\infty\;\|B_t\|_{_1}\;t^m\hat{\psi}(t) \;dt
=\int_0^\infty\;\|B_t\|_{_1}\;
t^m\hat{\psi}(t)\;dt.
\end{eqnarray*}
Let us estimate $\|B_t\|_{_1}$. For  $t>\epsilon$ with a fixed arbitrary small $\epsilon$, we have:
$$
\|B_t\|_{_1}=\|L^\Th(g)e^{-t\,\triangle_r^\Th/2\Lambda^2}\|_{_2}^2\leq
\|L^\Th(g)e^{-\epsilon \,\triangle_r^\Th/2\Lambda^2}\|_{_2}^2 \;
\|e^{-(t-\epsilon)\,\triangle_r^\Th/2\Lambda^2}\|.
$$
But, $(t-\epsilon)\triangle_r^\Th$ being positive, we have $\|e^{-(t-\epsilon)\,\triangle_r^\Th/2\Lambda^2}\|\leq1$.
Hence for $t>\epsilon$, $\|B_t\|_{_1}\leq C$ uniformly in $t$.
For $t\leq\epsilon$, our previous computation shows that $\|B_t\|_{_1}=O(t^{-n/2})$. Hence
$\Tr\left( \int_0^\infty B_t \;t^m\hat{\psi}(t)\;dt \right) <\infty$, so by dominated
convergence one obtains:
\begin{eqnarray*}
&&\hspace{-1.3cm}\Tr\left(L^\Th(\rho)\;\phi(\triangle_r^\Th/\Lambda^2)\right) \\
&=&(-1)^m\int_0^\infty\Tr\left(L^\Th(\rho) e^{-t\,\triangle_r^\Th/\Lambda^2}\right)t^m\hat{\psi}(t) \; dt\\
&=&(-1)^m(4\pi)^{-n/2}\int_0^\infty\sum_{l=0}^{m}\Lambda^{n-2l}\;t^{m+l-n/2}
\hat{\psi}(t) \; dt\;\int_{\R^n}\rho\Mop\;\tilde{a}_{2l}(x)\; d^nx \; +\;O(\Lambda^{n-2(m+1)})\\
&=&(4\pi)^{-n/2}\;\sum_{l=0}^{m}\Lambda^{n-2l}\,\phi_{2l}
\int_{\R^n}\rho\Mop\;\tilde{a}_{2l}(x)\;d^nx
\;+\;O(\Lambda^{n-2(m+1)}),\\
\end{eqnarray*}
where $\phi_{2l}$ is now defined by
\begin{equation}
\label{moments2}
\phi_{2l}:=(-1)^m\int_0^\infty t^{m+l-n/2}\,\hat{\psi}(t) \; dt.
\end{equation}
When $n=2m$ is even, $\phi_{2l}$ has the more familiar form of (\ref{moments}):
\begin{eqnarray}
\label{moments3}
\phi_{2l}=
\begin{cases}
{\textstyle{\frac{1}{\Gamma(m-l)}}}\;\int_0^\infty \phi(t) \; t^{m-1-l} \; dt,\;& {\rm for}\;
l=0,\cdots,m-1,\cr
(-1)^{l}\;\phi^{(l-m)}(0),& {\rm for}\; l=m,\cdots,n.
\end{cases}
\end{eqnarray}
For $n$ odd, the coefficients
$\phi_{2l}$ have less explicit forms because they invoke fractional derivatives of $\phi$,
so in this case, it is better to stick to definition (\ref{moments2}).

Let us summarize:

\begin{thm}
Let $\rho\in\SS(\R^n)$, $A=-iL^\Th(A_\mu)\otimes \gamma^\mu$,
$A_\mu^*=-A_\mu\in\Oh_0(\R^n)$, $\phi\in\mathcal{C}^\infty(\R^+)$
be a positive function satisfying condition (\ref{condition}) and
$\dslash_A=\dslash +A$. Then $L^\Th(\rho)\,\phi(\dslash_A^2/\Lambda^2)$
is trace-class. Moreover, the following expansion of the spectral action holds:
$$
S(\dslash,A,\rho)=2^m\;
 (4\pi)^{-n/2}\;\sum_{l=0}^{m}\Lambda^{n-2l}
\; \phi_{2l}\;\int_{\R^n} \rho(x)\;\tilde{a}_{2l}(x) \; d^nx
\;+\;O(\Lambda^{n-2(m+1)}),
$$
where the $\phi_{2l}$ are defined in (\ref{moments2}) or
(\ref{moments3}) depending on the dimension and the
$\tilde{a}_{2l}(x)$  are given in Theorem \ref{HKNCP}
with the following replacement in (\ref{genelap}):
$$
\begin{cases}
L^\Th(\omega_\mu) & \rightarrow \;L^\Th(A_\mu),\\
L^\Th(E) \otimes 1_{2^m}& \rightarrow \; \left(L^\Th(\pa_\mu A_\nu)
+L^\Th(A_\mu\Mop A_\nu)\right)
\otimes\frac{1}{2}(\gamma^\mu\gamma^\nu-\gamma^\nu\gamma^\mu ).
\end{cases}
$$
Moreover, all terms in
$\tilde{a}_{2l}$ linear in
$E$ are zero.
\end{thm}
\begin{proof}
This follows from $\gamma^{\mu}\gamma^{\nu}=\eta^{\mu \nu} +
{\textstyle{\frac{1}{2}}}(\gamma^{\mu}\gamma^{\nu}-\gamma^{\nu}\gamma^{\mu})$,
so all linear terms in $E$
are of zero trace.
\end{proof}

\begin{rem}
When the Dirac operator is symmetrized, $\D_A=\D+A+\epsilon JAJ^{-1}$, one has to
replace $L^\th(A_\mu)$ by $L^\th(A_\mu) - R^\th(A_\mu)$ since $\epsilon J\; (L^\th(A_\mu)
\otimes \ga^\mu)\;J^{-1}=R^\th(A_\mu^*) \otimes\ga^\mu$. So, the behaviour
in $t$ of different terms like $\Tr \big(L^\th(f)R^\th(g)\pa^\alpha e^{t\pa_\mu
\pa^\mu }\big)$ has to be computed. Since
$\sigma[L^\th(f)R^\th(g)](\xi,x)=f(x-\tfrac{1}{2}\th\xi)\,
g(x+\tfrac{1}{2}\th\xi)$, the translation invariance $x \rightarrow
x+\tfrac{1}{2}\th\xi$ crucially used in the proof of Theorem \ref{HKNCP} now fails.
This point is related to the UV{\tt /}$\,$IR mixings and has to be clarified.
\end{rem}

\subsection{Connes--Lott versus Chamseddine--Connes actions}\label{comparaison}

In order to compare this result with the Connes--Lott
action computation of the four dimensional
Moyal plane\cite{Gayral}, up to negative order terms with respect to the mass scale
$\Lambda$, we obtain:
$$
S(\dslash,A,\rho)={\textstyle{\frac{1}{4\pi^2}}}\left(\Lambda^4\phi_0\;
\int_{\R^4} \rho(x) \; d^4x+{\textstyle{\frac{\phi(0)}{6}}}
\;\int_{\R^4} \rho(x) \;  F^{\mu\nu}
\Mop F_{\mu\nu}(x) \; d^4x\right) +O(\Lambda^{-2}),
$$
where $F^{\mu\nu}:=\pa^\mu A^\nu-\pa^\nu A^\mu
+[A^\mu,A^\nu]_{\Mop}$.

If we choose the characteristic
function $\rho=\chi_{_V}$ of a bounded subset $V\subset \R^4$ ($\chi_{_V}\notin\SS(\R^4)$),
property (\ref{trace}) yields:
\begin{equation}
\label{actionspect}
S(\dslash,A,\chi_{_V})={\textstyle{\frac{1}{4\pi^2}}}\left(\Lambda^4\,\phi_0\;
\int_{V}d^4x+{\textstyle{\frac{\phi(0)}{6}}}
\;\int_{V} F^{\mu\nu}
\Mop F_{\mu\nu}(x) \; d^4x \right ) + O(\Lambda^{-2}),
\end{equation}
which, modulo a cosmological term, is the spatially localized
noncommutative Yang--Mills action. This expression has to be compared with the four
dimensional Connes--Lott one (\ref{NCYM}) for $\Th$ symplectic, hence $\Mop=\mop$:
$$
YM(\alpha)={\textstyle -\frac{1}{4g^2}}\int F^{\mu\nu}\mop F_{\mu\nu}(x)\;d^4x.
$$
This action is slightly different from (\ref{actionspect}), because property (\ref{trace}),
together with the absence of $\rho$ gives
$$
YM(\alpha)={\textstyle -\frac{1}{4g^2}}\int F^{\mu\nu}(x)F_{\mu\nu}(x)\;d^4x.
$$

For the noncommutative tori, we have also a similar result,
with the spectral action in the unital case ($\rho=1$):
$$
S(\dslash,A)\sim_{\Lambda \rightarrow \infty} \; 2^m \;(4\pi)^{-n/2} \; \sum_{k\in\N}\Lambda^{n-2k}
\; \phi_{2k}\;\int_{\T^n} \tilde{a}_{2k}(x) \; d^nx ,
$$
which also yields for $n=4$:
$$
S(\dslash,A)={\textstyle{\frac{1}{4\pi^2}}}\left(\Lambda^4 \, \phi_0\;
+{\textstyle{\frac{\phi(0)}{6}}}
\;\int_{\T^4} F^{\mu\nu}
\Mop F_{\mu\nu}(x) \; d^4x \right ) + O(\Lambda^{-2}).
$$

\subsection{Towards a gravitational degree of freedom}
One can ask about adding gravitational degrees of freedom for Moyal planes
or NC tori. For instance, the results of Section 3 also work with a non constant
metric $g^{\mu\nu}(x)$. More precisely, Theorem \ref{traceclass} is still true
if we replace in (\ref{genelap}), $\triangle^\Th$ by the square of
$$
-ie_a^\mu \big(\pa_\mu+\omega_\mu + L^\Th(A_\mu)\big)\otimes \gamma^a,
$$
where $e_a^\mu$ and $\omega_{\mu}$ are bounded functions. Here, the poinwise and Moyal
products are mixed, but in this case, computation of the trace of its regularized semigroup can be done, at least
in principle, with the same techniques but it will be highly less easy.

However, this construction is meaningless from a spectral triple point of view.
A non flat Dirac operator over $\R^n$, $\Dslash=-ie^\mu_a(x)\gamma^a(\pa_\mu+\omega_\mu(x))$,
$\omega_\mu$ being the spin connection, will violate most of the axioms describing
spectral triples, for instance
$$
[\Dslash,\pi^\Th(f)]=-i\gamma^a\left([e^\mu_a\,\omega_\mu,\pi^\Th(f)]
+e^\mu_a\,\pi^\Th(\pa_\mu f)+[e^\mu_a,\pi^\Th(f)]\,\pa_\mu\right).
$$
So, for $f\in\B_\Th$, $[\Dslash,\pi^\Th(f)]$ can be extended to a bounded operator only if
$[e^\mu_a,\pi^\Th(f)]=0$.
This condition can be satisfied for instance by a $n$-dimensional
Riemannian manifold $(M,g)$ endowed with an isometric action of $\R^l$, $l\geq 2$ (periodic or not).
This is actually the Connes--Landi isospectral deformations
\cite{ConnesDubois,ConnesLandi}. Those cases, admit nontrivial fluctuations of the metric (in some
sense for the untwisted directions). Since the only invariant metric on $\R^n$ or $\T^n$ by
the natural action of $\R^n$ is the flat one and this is the geometrical obstruction to deal with
nonflat Moyal planes (see \cite{ChamseddineGrav}).

\vspace{1\baselineskip}
\subsection*{Acknowledgments}
\hspace{\parindent}
It is a pleasure to thank J. M. Gracia-Bond\'{\i}a, T. Sch\"ucker, J. C.
V\'arilly, D.V. Vassilevich and V. Zagrebnov for fruitful discussions.

\end{document}